\input epsf
\documentclass[nohyper,notoc]{article} 
\usepackage{epsfig}
\usepackage{bm}

\def\beq{\begin{equation}}
\def\eeq{\end{equation}}
\def\bea{\begin{eqnarray}}
\def\eea{\end{eqnarray}}
\def\bq{\begin{quote}}
\def\eq{\end{quote}}
\def \lsim{\mathrel{\vcenter
     {\hbox{$<$}\nointerlineskip\hbox{$\sim$}}}}
\def \gsim{\mathrel{\vcenter
     {\hbox{$>$}\nointerlineskip\hbox{$\sim$}}}}
\def\gappeq{\mathrel{\rlap {\raise.5ex\hbox{$>$}}
{\lower.5ex\hbox{$\sim$}}}}
\def\lappeq{\mathrel{\rlap{\raise.5ex\hbox{$<$}}
{\lower.5ex\hbox{$\sim$}}}}

\def\ETm{E_T \! \!  \! \! \! \! \!  /~~}

\def\pslash{ \, p   \! \! \! / ~ }

\def\m3e{\mu \to e \bar{e} e}

\def\m{\mu}

\begin{document}
\vspace*{-1in}
\renewcommand{\thefootnote}{\fnsymbol{footnote}}
\vskip 5pt
\begin{center}
{\Large {\bf 
Leptoquarks decaying to a top quark and a charged lepton at hadron colliders }}
\vskip 25pt
{\bf   Sacha Davidson $^{1,}$\footnote{E-mail address:
s.davidson@ipnl.in2p3.fr}, 
 and Patrice Verdier $^{1,}$}\footnote{E-mail address:
verdier@ipnl.in2p3.fr} 
 
\vskip 10pt  
$^1${\it IPNL, Universit\'e de Lyon, Universit\'e Lyon 1, CNRS/IN2P3, 4 rue E. Fermi 69622 Villeurbanne cedex, France }\\
\vskip 20pt
{\bf Abstract}
\end{center}

\begin{quotation}
  {\noindent\small 
We study  the sensitivity of the Tevatron and the 7 TeV
LHC to a leptoquark $S$ coupling to a top quark and a charged lepton
$L$ ($ = e$, $\mu$, or  $\tau$). 
For the Tevatron, we focus on the case $m_S < m_t$, 
where the leptoquark pair production cross section is large, 
and the decay is three-body: $S\to W b L^{\pm}$.
We argue that existing  Tevatron observations could
exclude $m_S \lsim 160$ GeV.
For $m_S > m_t$, we show that the LHC experiments with low 
integrated luminosity could be sensitive to such leptoquarks decaying to
$tl^{\pm}$ with $l=\mu$ or $\tau$.

\vskip 10pt
\noindent
}

\end{quotation}

\vskip 20pt  

\setcounter{footnote}{0}
\renewcommand{\thefootnote}{\arabic{footnote}}

\section{Introduction and Review}
\label{intro}

Leptoquarks \cite{LQrev} are bosons  which couple to  a lepton and a quark.  
Although they are
not known to address current issues in particle physics
(such as the identity of dark matter or the hierarchy
problem), they can be motivated in several ways.  Most pragmatically,
they are strongly interacting  and their decay  products
include leptons,
so they are interesting search candidates for hadron colliders.
The Tevatron sets bounds on leptoquarks which decay to
first and second  generation fermions, and to $b$s; in
this note, we consider  leptoquarks which couple
to the top quark and any charged lepton $L^{\pm}$
($L \in \{ e$, $\mu$, $\tau \}$). We discuss the bounds that could be set with
4.3 $fb^{-1}$ of Tevatron data, and the prospects for
the 7 TeV LHC with 1 $fb^{-1}$.

Leptoquarks can arise in several extensions of
the Standard Model, such as Grand Unified Theories \cite{pdec},
Technicolour \cite{Farhi:1980xs} and 
$R$-parity violating Supersymmetry\cite{Barbier:2004ez}.
We focus on scalar
leptoquarks called $S$, with   baryon  and lepton 
 number  conserving
interactions,  and  a mass  $m_S \lsim 1$ TeV. 
 Several recent models 
\cite{Gershtein:1999gp,BG,models} include
such leptoquarks.
The Lagrangian describing  their renormalisable interactions
with  Standard Model (SM)
fermions and singlet neutrinos $\nu$ is \cite{BRW}
\bea
{\cal L}_{LQ} & = & 
S_0 ( {\bf \lambda_{L S_0}} \overline{\ell} i \tau_2 q^c
+ {\bf \lambda_{R S_0}} \overline{e} u^c ) +
\tilde{S}_0 {\bf \tilde{\lambda}_{R \tilde{S}_0}} 
\overline{e} d^c   
+
 ( {\bf \lambda_{L S_2}} \overline{\ell}  u
+ {\bf \lambda_{R S_2}}  
\overline{e } q [i \tau_2  ])S_{2} +
 {\bf \tilde{\lambda}_{L \tilde{S}_2}} 
\overline{\ell} d \tilde{S}_2
+  \overline{\ell} [i \tau_2  ] \vec{\tau} q^c \cdot \vec{S}_3 
 \nonumber \\
 &  & 
+ S'_0 {\bf \lambda'_{R S_0}} \overline{\nu} u^c  +
\tilde{S}'_0 {\bf \tilde{\lambda}'_{R \tilde{S}_0}} 
\overline{\nu} d^c   
+  {\bf \lambda'_{R S_2}}  
\overline{\nu } q [i \tau_2  ]S_{2} 
+ h.c. 
\label{BRW}
\eea
where the  leptoquark subscript
is its SU(2) representation,
 the $\lambda$s are 3 $\times$ 3 matrices
in the lepton and quark flavour spaces  and are labelled
by the SU(2) representation of the leptons ($L =$ doublet,
$R=$ singlet) and the leptoquark name, 
   $\tau_2$ is a Pauli
matrix  (so $i \tau_2$ provides the antisymmetric SU(2) contraction),
the SM SU(2)  singlets
are $e,u,d$ and $\nu$, and 
in this equation, $q$ and $\ell$ are the doublets.
For most of the rest of the paper, $L$ and $\ell$ label
physical particles
$$\ell \in \{ e , \mu\},~~~~~~~
L \in \{ e,\mu, \tau\}$$
In eqn (\ref{BRW}), we included  for completeness, leptoquarks which
couple to singlet neutrinos $\nu_R$. If the  neutrino masses
are Dirac,   these interactions  could allow
$S \to t \nu$ without $S \to b \nu$. 
However, we do not analyse such decays.
 Notice that we neglect,
or set to zero, 
the (renormalisable) interactions of the leptoquark
with the Higgs, which naturally
should be present, and can contribute via
loops to precision electroweak parameters \cite{STU}
and  neutrino masses
\cite{Hirsch:1996qy}.

\begin{figure}[ht]
\unitlength.5mm
\begin{center}
\epsfig{file=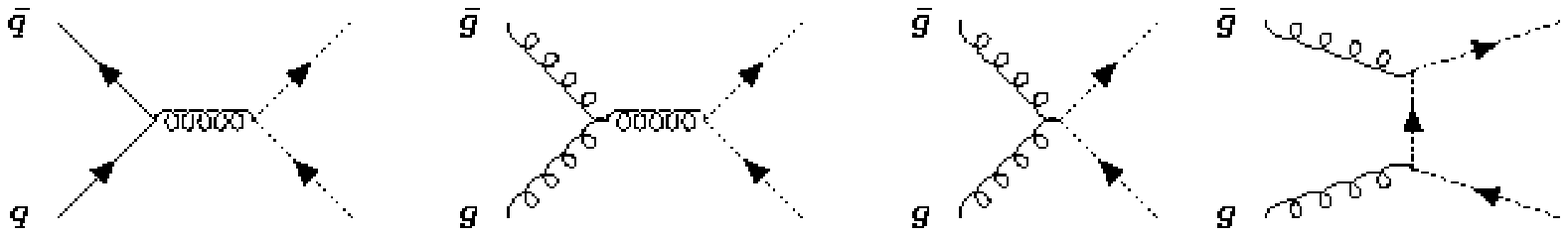,height=2.5cm,width=11cm}
\hspace{2cm}
\epsfig{file=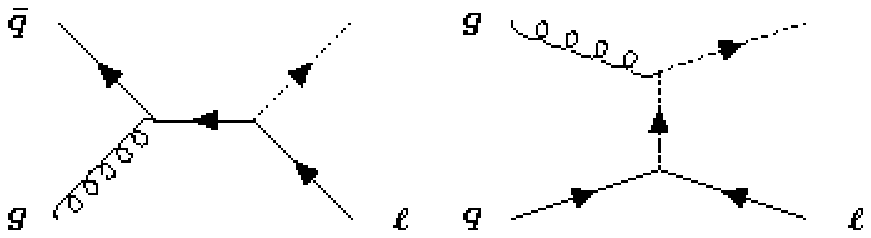,height=2.5cm,width=5cm}
\end{center}
\caption{ Lowest order diagrams for leptoquark
single and pair production;
single production can be neglected for a 
leptoquark which couples only to $t$ quarks,
because the $t$ content of the proton is small.
\label{prod1} }
\end{figure}

To look for leptoquarks, some theoretical
expectations about the structure and hierarchy of their
interactions would be helpful. 
Various theoretical arguments can suggest
 that the  largest leptoquark  couplings
should be to the third generation,
at least in the quark sector. This arises,
for instance, in the  Cheng-Sher ansatz \cite{CS}
for flavoured couplings
\beq
\lambda^{LQ} \propto \sqrt{\frac{m_L m_Q}{v^2}}
\eeq
where $ v = 175$ GeV is the Higgs vacuum expectation value.
This would give the largest leptoquark coupling
to $t$ and $\tau$,  and can arise Randall-Sundrum
type extra dimensional models, or in composite
models as recently discussed  in \cite{BG}. 
The expectations of this ansatz are compared to
current low energy constraints in \cite{Carpentier};
improving the sensitivity of $K \to \pi \nu \bar\nu$ 
could probe this pattern for  leptoquarks that
couple to neutrinos.

\begin{figure}[ht]
\begin{center}
\epsfig{file=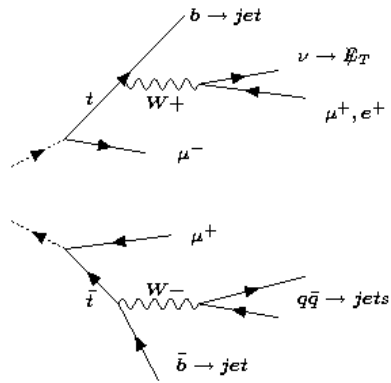,height=6cm,width=8cm}
\end{center}
\vspace{-10mm}
\caption{Possible decay chain for a pair of scalar
leptoquarks interacting with tops and muons. 
}
\label{fig:etatfin}
\end{figure}

A phenomenological ``bottom-up'' approach to  
the couplings of new particles,
 motivated
by the success of the Minimal Flavour Violation
hypothesis \cite{MFV,dAGIS}, 
is to construct them by multiplying
  the known mass  and mixing matrices
of leptons and quarks. Some
possibilities for leptoquarks were
studied in \cite{MFVLQ}.
In this approach,
Nikolidakis and Smith \cite{NS}
noted that  a New Physics
coupling with a single  lepton index $L$ can be
proportional to
\beq
\varepsilon^{LJK} [Y_e Y_e^\dagger m_\nu]_{JK}
\label{CS}
\eeq
where $Y_e$ is the charged lepton Yukawa matrix
(index order doublet-singlet), and $m_\nu$ is
the majorana neutrino mass matrix. This
idea was studied for leptoquarks in \cite{MFVLQ}. 
Since $m_\nu$ is fairly democratic, the
$Y_eY_e^\dagger$ hierarchy selects the $\tau$ index.
The totally antisymmetric SU(2) tensor is $\varepsilon$, 
so this construction favours couplings to the
$e$ and $\mu$. It is therefore interesting to study
leptoquarks which decay to $t$ and any charged  lepton:
$L = \tau$, $\mu$, or  $e$. The $\mu$ and $e$
are  particularily attractive final state particles  
for hadron colliders:
if one of the $W$s from  the $t$s decays to the
$\ell = e$ or $\mu$, as occurs $\sim$ 29 \%  of
the time, the final state would be
jets $ + \ETm + \ell^\pm \ell^\pm \ell^\mp$
(see figure \ref{fig:etatfin}.)

\begin{figure}[ht]
\begin{center}
\epsfig{file=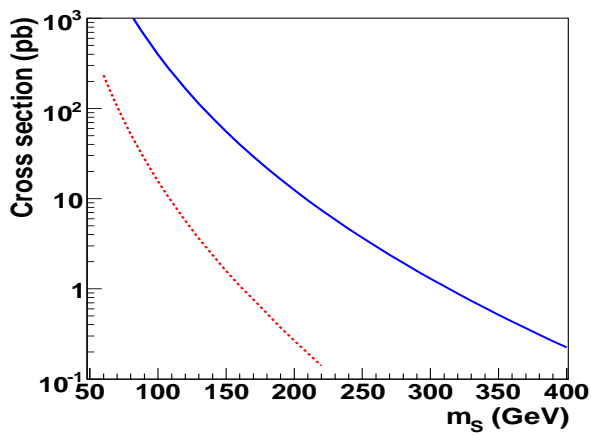,height=6cm,width=8cm}
\end{center}
\caption{Pair production cross section for SU(2) singlet 
leptoquarks at the Tevatron and the 7 TeV LHC.
 Standard Model
$t \bar{t}$ production, with  
$\sigma_{pp \to t \bar{t}}(\sqrt{s} = 7~{\rm TeV}) \simeq
165\, $pb  and 
 $\sigma_{p\bar{p} \to t \bar{t}}(\sqrt{s} = 1.96~{\rm TeV}) \simeq
7.8\, $pb, 
 could be a
significant background to the leptoquark
signal, in particular in the $m_S \sim m_t \pm 15 $ GeV
region, where the leptons leptoquark decays
would be soft. }
\label{fig:seceff}
\end{figure}

There are various experimental constraints on leptoquarks.
At hadron colliders, they 
 can be singly or pair produced 
via their strong interactions (see figure \ref{prod1}).
As discussed in \cite{Belyaev}, single production can lead to the
same  final state as pair production. However, 
since we are interested in
leptoquarks that couple  to the top quark, 
we can  neglect single production,   because
the top density in the proton is negligeable. 
 The
cross section, for   pair production
from $gg$ or $ q \bar{q}$, has been
computed at Next to Leading Order (NLO)~\cite{coll}, and included 
in the {\sc prospino} program, which we used to
produce figure~\ref{fig:seceff}.  
 The Tevatron
\cite{tevatron} has searched for
leptoquarks  decaying
to any lepton and a quark other than the top,
 with a coupling
$\lambda \gsim 10^{-8}$.
The restriction on $\lambda$   ensures that
the leptoquark decays at the collision point. 
The bounds depend on the final
state; a recent review \cite{gerald}
gives   $m_S \gsim 210 [\tau b],\ 
 214 [\nu q],\ 
247[\nu b],\ 299 [e q, e b ],\ 316 [\mu q ,
\mu b]$\,
GeV, where $q  \in \{ u,d,s,c\}$. 
Leptoquarks have also been searched for at the HERA $ep$ collider, 
which allow to exclude a $s$-channel resonance with $\lambda \gsim 0.1$ 
and $m_S \lsim 250 - 300$\,GeV \cite{HERA}.
Finally, there are  bounds  
on two quark, two lepton contact interactions
from several (mostly accelerator) experiments \cite{Lambda}, which
give interesting constraints
(see {\it e.g.} \cite{Carpentier},
\cite{Cheung}) on 
leptoquarks interacting
with lower generation fermions. 
The prospects of discovering leptoquarks
above the various backgrounds at the early LHC
have been discussed in \cite{bckgrds}.

There are numerous  precision/rare decay bounds  on
leptoquarks, which usually  apply to products of different  $\lambda$s,
and depend on the SU(2) representation of the leptoquark.
Some recent compilations are \cite{PRD} (bounds from meson
anti-meson mixing, allowing for complex couplings)
and \cite{Carpentier} (mostly tree processes). 
In general, it is clear that these bounds exclude  
flavour-democratic  $\lambda^{LQ} \sim {\cal O}(1)$ 
for  $m_S < $ TeV. Certain processes, such
as $K \to \pi \nu \bar\nu$ or $K_L \to \mu^\pm e^\mp$
provide much more stringent bounds. 

Bounds and prospects for a ``third generation''
 leptoquark  interacting
with a top, have been discussed by several  people,
in particular Eboli and collaborators.
 The constraints   from the loop
contribution to leptonic
$Z$ decay  \cite{Concha} are satisfied
 if    $\lambda \lsim e$ for $m_S \sim 300 $ GeV
(for both   the leptoquarks of eqn (\ref{BRW})  which
couple to $t_R$). To extrapolate
this bound for  leptoquarks in the range
300 GeV $\to m_W$, we assume that  the bound  can
be scaled
as $\lambda/m_S \lsim e/(300 GeV)$, see eqn
(\ref{thdreams}).
The LHC prospects of a leptoquark
decaying to $t \tau$ or $b \tau$ were discussed in
 \cite{Ebolill}, who emphasized the interesting 
one, two and three lepton final states
which could be detected above backgrounds.
Gripaios et al recently  implemented 
the various leptoquarks of eqn(\ref{BRW}) 
in {\sc herwig}~\cite{BG2}, and discussed kinematic
reconstruction techniques for  
leptoquarks  decaying to third
generation fermions at the 7 TeV LHC. 


In this paper, we study 
leptoquarks which couple to tops, but not
to $b$s or lower generation quarks,  
because  leptoquarks with
an ${\cal O}(1)$ branching ratio to
$b \nu$, or jet  + $ e$ or $\mu$,
 are  already excluded  by
the Tevatron up to $m_S \lsim 200-300$ GeV  \cite{tevatron,gerald}.
We are therefore interested in
 leptoquarks which couple to singlet
up-type quarks,  that is,  
the  SU(2) singlet leptoquark $S_0$
with coupling $\lambda_{R S_0}$ (and
 $\lambda_{L S_0} = 0$), or
the doublet leptoquark $S_2$ with
couplings  $\lambda_{L S_2} \neq 0$ and
 $\lambda_{R S_2} = 0$. 
Neither of these leptoquarks
arise in $R$-parity violating
Supersymmetry.
We then restrict the coupling
to third generation quarks, and 
assume, for the body of the paper,  a
branching ratio  of 1 to the final
state of top +  the charged
lepton $L^{\pm}$  of our choice.

The NLO cross section for leptoquark
pair  production~\cite{coll}, via the strong interaction, 
is plotted in figure~\ref{fig:seceff}. This
shows  that the  Tevatron with 5 $fb^{-1}$
of data could produce  $ \gsim {\cal O} (5000)$ 
pairs of   leptoquarks
with $m_S \lsim 150$ GeV, whereas the $7$ TeV
LHC with $1 fb^{-1}$ could  produce  a thousand pairs
of 300 GeV leptoquarks.
Section \ref{D0}  outlines  a simple
counting experiment that compares  current
Tevatron  data, to 
leptoquarks with $m_S < m_t$, decaying
via an off-shell top to $ b W$ and a charged lepton. 
It suggests that the Tevatron could exclude
such leptoquarks, for leptoquark masses
sufficiently below $m_t$.  
 In section
\ref{LHC}, we briefly mention prospectives with 1 $fb^{-1}$
of data from the 7 TeV LHC. 

\section{ Leptoquarks  with  $ \bm{m_S < m_t}$ at the Tevatron}
\label{D0}

In this section, we consider  leptoquarks $S$ interacting
only with the top (and any $L$),   with
masses  in the range $m_W+ m_b < m_S < 2 m_W < m_t$.
They  would be copiously
pair-produced  via their strong interactions
at the Tevatron. There are two reasons for
this limited mass  range,
despite that  figure (\ref{fig:seceff}) suggests
several hundred leptoquark pairs could be 
produced at the Tevatron  up to masses
$m_S \sim 250$ GeV.  Firstly, in the range
$m_S \simeq m_t \pm 15$ GeV, the lepton produced with the 
almost-on-shell
top is unlikely to pass the  $p_T > 15-20$ GeV cuts
that we impose.
Secondly, it is 
convenient to analyse separately the $m_S < m_t$
and $m_S > m_t$ cases; so we study
the former at the Tevatron and the latter at the LHC. 

Leptoquarks with $m_S < m_t$  could also be singly produced
in top decay; however we neglect this process,
because $BR(t \to SL) \leq 2|\lambda|^2/|y_t|^2$
is suppressed by the  leptoquark
coupling $\lambda$ ($y_t$ is the top quark Yukawa coupling).  
We consider the range 
\beq
10^{-(6\div 3)} < \lambda \ll e\frac{m_S}{300~{\rm GeV}}
\label{thdreams}
\eeq
where the upper bound 
is approximately the constraint from leptoquark
loop contributions to the $Z \bar{L} L$ vertex\cite{Concha}. 
It  implies that 
 $BR(t \to SL)$ is   negligeable,  when it is kinematically allowed. The lower 
bound ensures  that $\lambda$ is sufficiently large
that the leptoquarks decay at the collision
point (no displaced vertex);  see the discussion
after eqn (\ref{eq:f14}).  
We also assume  that $S$ interacts only with a top and
a charged lepton $L^{\pm}$, so  $BR(S \to b W L)$ = 1,
where $L$  is a $e, \mu,$ or $\tau$.

Since the leptoquark decays to a singlet $t_R$, 
its decay rate  to  $b W L$, via
an off-shell $t$,  has a simple analytic
form, which we obtain in subsection \ref{mS<mt}. 
This allows us to implement the three-body
decay in {\sc pythia}, as the product of the
two body decays $\Gamma( S \to t^* L) \Gamma(t^* \to b W)$
with variable $m_t$ and leptoquark coupling $\lambda$. 
This is discussed at the end of section  \ref{mS<mt}. 

Subsection \ref{D0bds} contains preliminary
estimates of the contribution of such leptoquarks
to the jets + $\ETm$ +  $\ell^{\pm}$  \cite{sum} and
jets + $\ETm$ +  $\ell_i^\pm
\ell_j^\mp$
\cite{win} data sets used by D0 to measure
the $t \bar{t}$ production cross section
($\ell$ here means $e$ or $\mu$). 
We consider separately the cases $S \to t \tau^\pm$
and $S \to t \mu^\pm$; we assume that  the bounds
which could be obtained on $S \to t e^\pm$  are  similar to
those on $S \to t \mu^\pm$.  

\subsection{The decay rate $\bm{S \to b W L^{\pm}}$ for $\bm{m_S < m_t}$}
\label{mS<mt}
\label{caln}


If the masses of the $b$  and $L \in \{ e, \mu ,\tau \} $ 
are neglected, then the invariant mass of the $bW$ system
(or equivalently, the magnitude of the four-momentum
carried by the off-shell top in the decay $S \to  bWL$),
is
\beq
t^2 = m_{bW}^2 = (p_b + p_W)^2 = 2 p_b \cdot p_W + m_W^2 ~~.
\eeq
The differential  three-body decay rate can be written
\cite{PDG} 
\beq
\frac{d \Gamma}{dt} = \frac{1}{(2 \pi)^5}\frac{1}{16 m_S^2}
\int
|{\cal M}|^2 |\vec{p}^{~*}_b| |\vec{p}_L| d \Omega^*_b d\Omega_L 
\label{dGdt}
\eeq
where  the $L$ parameters are in the $S$ rest frame,
and the starred $b$ parameters  in the $bW$ rest frame.

The matrix element for $S \to b WL$  is
\beq
{\cal M} =  \frac{\lambda g}{\sqrt{2}}
\overline{u}_{L} P_R \frac{\pslash_{\! \!  t} + m_t}{t^2 - m_t^2 + i m_t \Gamma_t}
\gamma^\mu P_L  u_b \varepsilon_\mu
\eeq
(where $u_L$ is the spinor field for $L$),
and is simple in squared form because the top
must flip chirality on the internal line:
\beq
|{\cal M}|^2 = - \frac{m_t^2 \lambda^2 g^2}{(t^2 - m_t^2)^2 + m_t^2 \Gamma_t^2}
\left( p_L \cdot p_b + 2 \frac{ p_L \cdot p_W   p_b \cdot p_W}{m_W^2}
\right) ~~.
\label{M2}
\eeq

To evaluate the angular integrations of eqn (\ref{dGdt})
with $|{\cal M}|^2$ from eqn (\ref{M2}), requires the Lorentz
transformation of the $b$  4-momentum 
in the $S$ frame ($p_b$), to the $t$ frame($p_b^*$). 
Writing 
\beq
E_b = \gamma E_b^*(1 + \beta \cos \theta^*)
\eeq 
with  $\gamma = E_t/m_t$, $\beta = |\vec{p_t}|/m_t$,
and using 
\beq
|\vec{p}_b^*| =  \frac{t}{2} \left( 1 - \frac{ m_W^2}{t^2} \right)
~~~~,~~~
|\vec{p}_L| =  \frac{m_S}{2} \left( 1 - \frac{ t^2}{m_S^2} \right) ~~,
\eeq
gives
\bea
  \int 
\left( p_L \cdot p_b + 2 \frac{ p_L \cdot p_W   p_b \cdot p_W}{m_W^2}
\right) 
 d \Omega^*_b d\Omega_L 
= 4 \pi^2 m_S^2
\left( 1 - \frac{t^2}{m_S^2} \right)
 \frac{t^2}{m_W^2}  
\left( 1 - \frac{m_W^2}{t^2} \right)
\left( 1  +2 \frac{m_W^2}{t^2} \right)
\label{true?}
\eea
So one obtains
\beq
\frac{d \Gamma}{dt} = 
\frac{\Gamma(S \to t^* L)}{2m_t}
\frac{\Gamma (t^* \to W b)}{\pi m_t} \frac{m_t^4}{(t^2 - m_t^2)^2 + m_t^2 \Gamma_t^2}
\label{dG}
\eeq
where the two body decay rates of the leptoquark $S$ and the
top quark $t$ are the rest-frame formulae,
with $m_t$ replaced by $ t$:
\beq
\Gamma (t^* \to W b) = \frac{g^2 t }{64 \pi} \frac{t^2}{m_W^2}  
\left( 1 - \frac{m_W^2}{t^2} \right)^2
\left( 1  +2 \frac{m_W^2}{t^2} \right)
~~~~~,~~~~
\Gamma(S \to t^* L) = \frac{\lambda^2 m_S}{16 \pi} 
\left( 1 - \frac{t^2}{m_S^2} \right)^2 ~~.
\eeq

\begin{figure}[ht]
\begin{center}
\epsfig{file=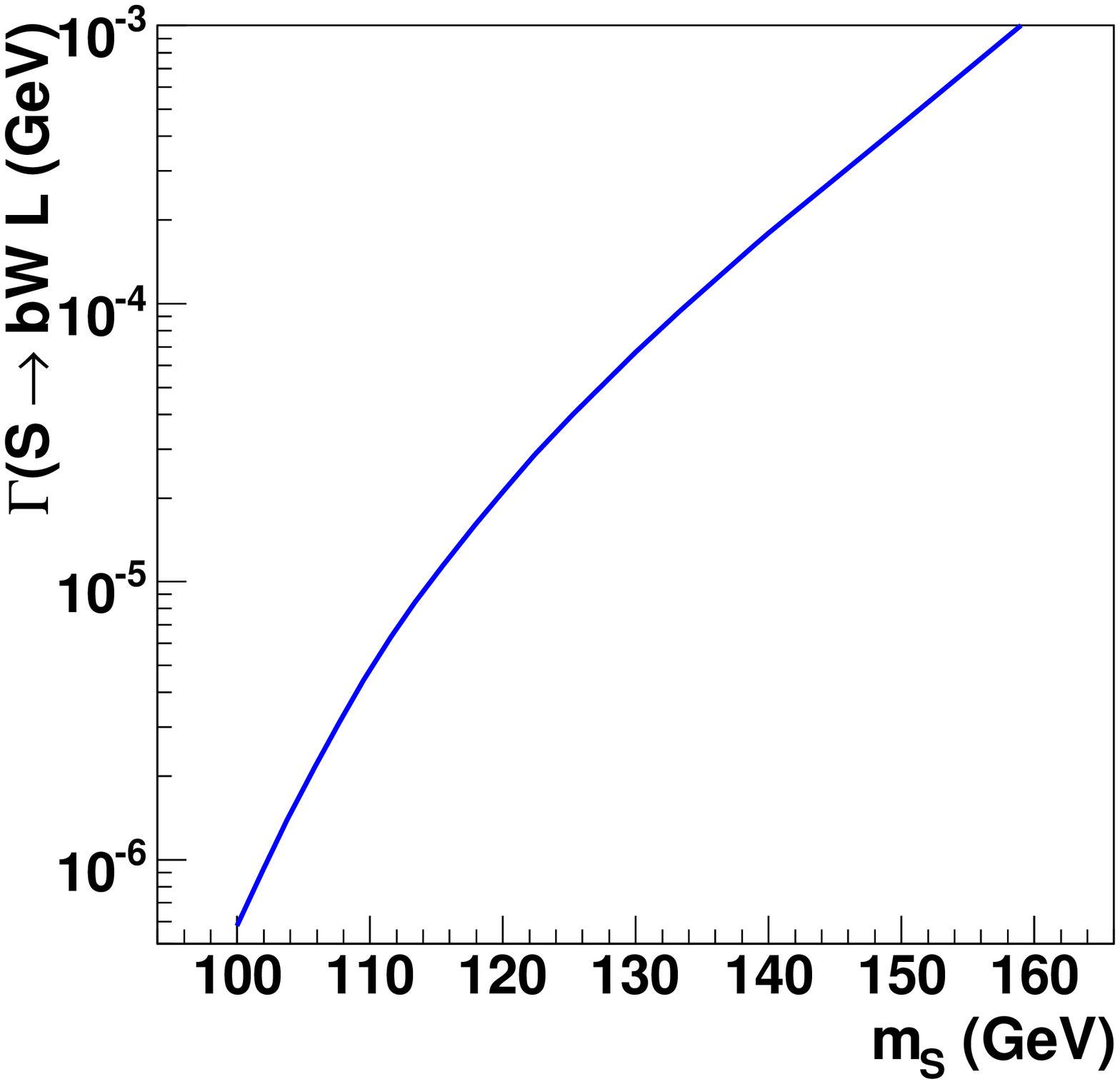,height=6cm,width=7cm}
\hspace{1cm}
\epsfig{file=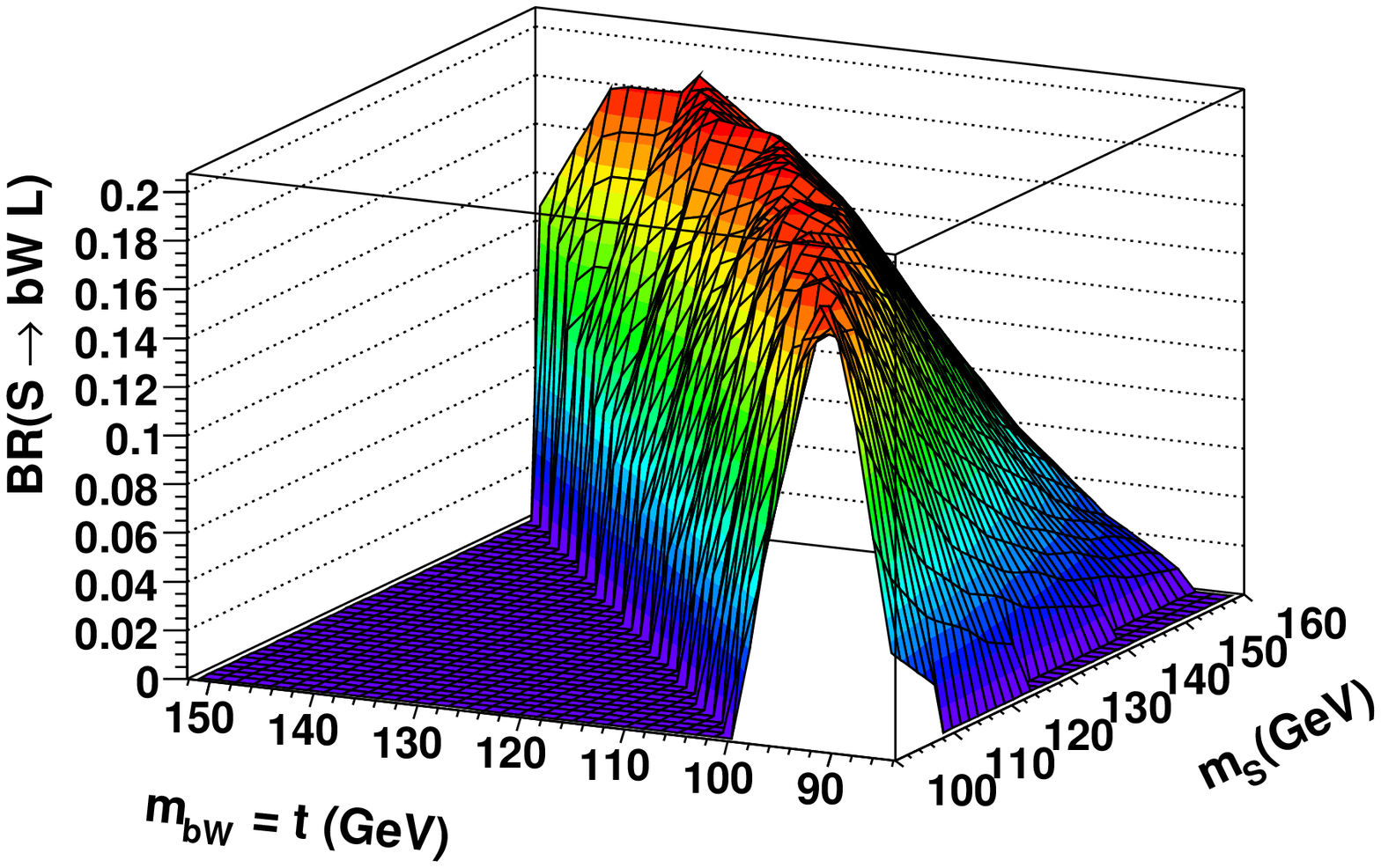,height=6cm,width=9cm}
\end{center}
\caption{  On the left, the total decay rate 
$\Gamma(S \to b W L)$, for $\lambda = 1$, as a function
of the leptoquark mass $m_S$, where $L$ is a charged
lepton $e, \mu$ or $\tau$.  On the right, the fraction
of decays to a $bW$ of invariant mass $t$ for various
leptoquark masses $m_S$.
}
\label{fig:Gam}
\end{figure}

A check  that can be performed on eqn(\ref{dG})
is to take the limit $t^2 \to m_t^2$.  Using the
 identity:
\beq
\frac{1}{\pi} \frac{\epsilon}{x^2 + \epsilon^2} \to \delta(x)
\label{eq:f14}
\eeq
with $x = t^2 - m_t^2$,
the $dt$ integration can be performed, and one obtains
the  two-body leptoquark decay
rate  $\Gamma(S \to t L)$, as expected.
We are interested in the case $t^2 - m_t^2 \gg \Gamma_t m_t$,
so we drop the $\Gamma_t$ term in the denominator of eqn (\ref{dG}).
The total decay rate $\Gamma(S \to bWL)$ can 
also  be obtained analytically, 
but is not illuminating. It is plotted on the left  in figure \ref{fig:Gam} 
for $\lambda = 1$. The leptoquark will decay in
less than a centimetre for $\lambda \gsim 10^{-3}$ at
$m_S \simeq 100$ GeV,  and for $\lambda \gsim ~{\rm few} \times 10^{-6}$
at $m_S \simeq 160$ GeV.

Equation (\ref{dG}) implies that 
 the decay $S \to bW L$ can be computed as
the two-body decay $S\to t^* L$, where
$t^*$ is a top quark of mass $t$,  followed
by the decay $t^* \to bW$, provided  the whole
process has a $t$-dependent coupling constant
$\propto  1/(t^2 - m_t^2)^2$. We implement
the $m_S < m_t$ leptoquarks in {\sc pythia} from
this perspective: for each event, the top
mass is randomly selected, distributed
between $m_W + m_b$ and $m_S$ according
to eqn (\ref{dG}).
The two-body leptoquark decay is then performed by {\sc pythia} 
followed by a two-body top decay. We checked that the
resulting $bW$ invariant mass distribution
reproduces  eqn(\ref{dG}). 

%

\subsection{Potential bounds from the Tevatron}
\label{D0bds}

In this section we estimate  the
 number of events from $m_S < m_t$ leptoquarks,
that could appear in two recent  D0 data samples
used to measure the $t \bar{t}$ production
cross section:
$\ell^{\pm} + \ETm $  + jets \cite{sum}, for
leptoquarks decaying to $t \tau^\pm$, and
$\ell^\pm_i \ell_j^\mp + \ETm $  + jets \cite{win}
for leptoquarks decaying to $t \mu^\pm$.
These two analyses are not optimal
to select these leptoquark signals because they 
require exactly one and two leptons, respectively,
in the final state. However, due to the important leptoquark
pair production cross section when $m_S < m_t$, we will show
that a significant number of leptoquark events would nevertheless
enter in these data samples.

We obtain the leptoquark events using {\sc pythia}\,6.4
\cite{pythia}
with {\sc tauola}~\cite{tauola} to decay $\tau$s\footnote{We modified
the {\sc tauola-pythia} interface
so that it finds and assigns polarisation 
to $\tau$s from leptoquark
decay.}, 
and reconstruct jets using the anti-$k_t$
algorithm of the {\sc fastjet} package~\cite{fastjet}.
All final state particles except neutrinos
and charged leptons are used 
to construct the jets. We do not include a detector
simulation. We only impose
the preselection cuts  of the experimental analyses
on our leptoquark events, and count
the number of remaining events.
D0 uses multivariate techniques 
to further discriminate  $t \bar{t}$ signal
from $W +$~jets or $Z +$~jets, which we do not consider. 
We study two cases: $S \to t \tau^-$, and
$S \to t \mu^+$. 

{In \sc pythia}, leptoquarks can decay to
$t$ and ${L}^+$, but not \footnote{The various scalar leptoquarks
of eqn (\ref{BRW}) have recently been included
in {\sc herwig}~\cite{BG2}, which would avoid
this limitation that {\sc pythia} only
knows one chiral structure for leptoquark
interactions.} 
to $t $ and $ L^-$. The first is
appropriate to the SU(2)
doublet leptoquark $S_2$, whereas
the second would correspond 
to the singlet leptoquark $S_0$.
In the $S \to t \mu$ case, the 
events corresponding to $t \mu^-$ or
$t \mu^+$ should be  equally detectable,
so we let {\sc pythia} decay the leptoquarks
to $ t \mu^+$, and apply the resulting
bounds to both $S_2$ and $S_0$. 

The case of $S \to t \tau^{\pm}$ is more
delicate, because the angular and
energy distribution of the tau decay
products depends on the charge and
polarisation of the $\tau$. We
therefore ask {\sc tauola} to flip
the sign of $\tau$s produced in $S_0$ decays.  
In the case of leptonic  $\tau$ decays, 
energetic $\ell^-$ are emitted
preferentially  anti-parrallel
to the direction of
motion of a relativistic $\tau_R$  (see {\it e.g.}
the $e$ distribution from $\mu$ decay in
\cite{PDG}).  So for the
singlet leptoquark $S_0$, which decays
to $t_R (\tau_R)^-$,  the neutrinos from
the leptonic tau decays frequently carry
most of the $p_T$ of the $\tau$. 
Since the $\tau$s are already not very
energetic, this means the charged
leptons from their decay may not pass $p_T$ cuts,
so hadronic $\tau$ decays are more
useful. We assume that the bounds
we obtain on the singlet leptoquark $S_0$
can be applied to  the doublet leptoquark $S_2$,
which decays $S_2 \to t (\tau_L)^+$, 
because  our bounds will mostly come from
events with hadronic $\tau $ decays\footnote{We checked 
that changing the $\tau$ polarisation makes 
a relatively  insignificant change to the bounds.}.

\subsubsection{$S \to t \tau ^\pm$} 
\label{sssttau}

Consider first  the production
of a leptoquark anti-leptoquark pair,  followed
by leptoquark decay to a $t$ and a  $\tau^-$ or $\tau^+$.  
This could contribute to the
$\ell^{\pm} + \ETm $  + jets signal from which D0 extracted 
the Standard Model top pair
production cross section  $\sigma_{t \bar{t}}$ 
in \cite{sum}; 
since  the leptoquark process
should have more jets than $t \bar{t}$ production,we
focus on the 
$\ell^{\pm}$ and at least $4$ jet sample. 

In our simulated sample of leptoquark events, 
we require that a  $W^\pm$ from  the $t$ or $\bar{t}$  decays
to $\nu$ and $e^\pm$ or $\mu^\pm$, 
which reduces the cross section by a factor
\beq
BR(\ell^\pm +  \ETm + n~jets) = .22 (W^+ \to \ell^+ \nu) 
\times .66(W^- \to had)  \times 2 (W^+ \leftrightarrow W^-)
\simeq .29
\label{BR14}
\eeq
Then we impose the following cuts, patterned on
the preselection of the D0 analysis.
We require
\begin{enumerate}
\item
 $\ETm > 25$ GeV
\item  a lepton with
 $p_T >20$ GeV and  $|\eta| \leq 1.1 (e),2.0(\mu)$  and
no second lepton with  $p_T >15$ GeV
\item  at least
 4 jets with  $p_T > 20$\,GeV, and  $|\eta| < 2.5$
\end{enumerate} 
With  $\varepsilon_{sim}$ the fraction of
simulated events which pass these cuts, the inclusive leptoquark signal efficiency
is simply $\varepsilon(\ETm, 1\ell, 4j)=\varepsilon_{sim} \times BR $. This efficiency
is given in colomn~3 of Table~\ref{tau3}.
We then estimate
the number of leptoquark induced events  in the 4.3 $fb^{-1}$
of data used in \cite{sum} to be 
%
the last colomn of table \ref{tau3}. 

\begin{table}[htp!]
\begin{center}
\begin{tabular}{|c|ccc|}
\hline
$m_S $ (GeV) &  $\sigma$ (pb) & $\varepsilon(\ETm, 1\ell, 4j)$ &$ N(LQ)$  \\
\hline
160 &  1   & .0823     &  367  \\
140 &  2.4 & .0618   &  658  \\
120 &  6   & .0389    & 1035  \\
100 &  16  & .0149  & 1060  \\
\hline
\end{tabular}
\caption{\label{tau3}The second colomn
is the leptoquark pair production cross section at the
Tevatron  for various
masses. The leptoquarks decay to $t \tau^\pm$.
The  third  colomn
estimates  the fraction of  events remaining after
the  cuts given 
in section \ref{sssttau}. The last colomn is
the expected number of leptoquark-induced
events in the D0 lepton +  $\geq 4$ jets sample \cite{sum}
based on 4.3 $fb^{-1}$.}
\end{center}
\end{table}

The total number of observed [expected]  $\ell^{\pm} + \ETm + \geq 4$~jets
events in \cite{sum} is $1795 [1796 \pm 158]$. Using the modified frequentist $CLs$
method
\cite{CL}
, the number of signal events excluded at 95\,\% C.L. is 388. By interpolating 
our results, we find that  leptoquarks with mass 
$$m_S < 158\, {\rm GeV}
~~~~{\rm  for} ~~~~~ BR(S \to t \tau^\pm) = 1
$$ are excluded at 95\,\%~C.L.

\subsubsection{$S \to t \mu ^\pm$} 
\label{mmmu}


Consider now a pair of leptoquarks which
decay to a $t$ and a $\mu^\pm$, which could
contribute to the $\ell^+_i \ell_j^- + \ETm$ + jets
data sample from which D0 extracts
$\sigma_{t \bar{t}}$~\cite{win}. In this analysis
\cite{win},  $\ell^+_i \ell_j^-$  can be 
$e^+ e^-$,$e^\pm \mu^\mp$ or $\mu^+ \mu^-$. 
The leptoquarks we study 
would contribute mostly to 
$e^\pm \mu^\mp$ or $\mu^+ \mu^-$. However,
to be conservative, we compare our expectations
to the observed number of  $\ell^+_i \ell_j^- + jets$ 
 events, including $e^+ e^-$. 

We simulate leptoquark pair production, followed
by  leptoquark decays $S \to W^+b\mu^+$, and require
that one $W$ decay to a charged lepton $e$, $\mu$ or $\tau$,
which should represent a fraction 
\beq
 .34 (W^+ \to \ell^+ \nu) 
\times .66(W^- \to had)  \times 2 (W^+ \leftrightarrow W^-)
\simeq .45
\label{BR15}
\eeq
of the events. We require a leptonic $W$ to ensure missing transverse energy
in the event, but  include also the $W \to \tau \nu$ decays, because
a lepton from the $W$ is not neccessary since two charged leptons are already
coming directly from the leptoquark decay.

Our cuts to select two charged leptons are patterned on the D0 analysis \cite{win}.
We therefore require exactly two opposite sign leptons of  $p_T > 15$ GeV
with $|\eta| \leq 1.1 (e),2.0(\mu)$. Then, D\O\ uses a multivariate technique
(Bayesian decision tree or BDT) to further discriminate top pair events from 
$Z/\gamma^{*}+$jets events which we can not take into account. But since the 
topology of our leptoquark signal is very close from the one of 
top pair production,
we believe that leptoquark events will nevertheless pass the 
selection cut applied 
on this BDT output with an efficiency very close to the
 one from top pair events.
In the following, we will not take into account the 
efficiency of the BDT, and simply
replace it by a cut $\ETm >25$ GeV. Then, we count 
the number of jets satisfaying 
$p_T >20$ GeV and $\eta<2.5$ and require at least 3 jets.

In  the third and fourth 
colomns of table \ref{tmu2}, we
give the estimated fraction of
leptoquark events which would pass
the above cuts (obtained
by  multiplying eqn (\ref{BR15}), 
and the fraction of simulated events
which pass cuts) and the expected
 number of leptoquark-induced
events in 4.3 $fb^{-1}$.
These numbers can be
compared to the  $\sim 51 [65 \pm 15]$  
observed[expected] events~\footnote{The numbers are
extracted from a histogram.},
bearing in mind that we have not simulated detector 
effets and that we did not 
take into account the efficiency of the selection cut on the 
BDT output rejecting 
Drell-Yan events, which is around 70\%.
From those events, we computed that the number of signal events 
excluded at 95\,\% C.L. is 39.
We therefore see that the expected number of leptoquarks 
events is much larger than this number:
this leptoquark signal would significantly contribute to the number of events observed in the 
D\O\ analysis, and we can conclude that leptoquark with mass
$$m_S < 160 ~ {\rm GeV}
~~~~~{\rm  for} ~~~~  BR(S \to t \mu^\pm) = 1
$$ are excluded at 95\%\,C.L.

\begin{table}[htp!]
\begin{center}
\begin{tabular}{|c|ccc|}
\hline
$m_S$ (GeV)&  $\sigma$ (pb) & $\varepsilon(\ETm, 2OS\ell,3j)$ & $N(LQ)$ \\ \hline
160 &  1  & .0900    & 387  \\
140 &  2.4  &.0752     & 776 \\
120 &  6  & .0500   & 1288 \\
100 &  16  &.0090   & 960\\
\hline
\end{tabular}
\caption{The second colomn
is the leptoquark pair production cross section for various
masses. The leptoquarks decay to $t \mu^\pm$.
The   third  colomn
estimates the fraction of  events remaining after
the  cuts given 
in section \ref{sssttau}. The last colomn is
the expected number of leptoquark-induced
events in the D0 $\ell^+ \ell^- +  \geq 3$ 
jets sample \cite{win} of 4.3 $fb^{-1}$. \label{tmu2}}
\end{center}
\end{table}


\section{At the 7 TeV  LHC}
\label{LHC}

Leptoquarks decaying to a top and an $e$ or $\mu$ are attractive
search candidates for the early LHC because the final state contains
leptons and many jets. If a $W^\pm$ from the $t$ or $\bar{t}$ decays
leptonically, various combinations of same sign and opposite sign
leptons of different flavour can be obtained (see figure 
\ref{fig:etatfin}). 

Since the events contain many jets, the leptoquark
pair production and decay should be calculated at
NLO, so that the Monte Carlo simulation matches
as well as possible to the real events.
 In addition,  detailed study of backgrounds
would be required to identify suitable  cuts
to select leptoquark events and identify
the leptoquark mass. We leave this analysis to
the experimental collaborations, and here, we merely
estimate the fraction of events  at the LHC,
with  cuts similar to recent LHC $t \bar{t}$ results \cite{CMS,ATLAS}.

We consider leptoquarks  that  decay with a branching ratio of
one to either $t \tau^-$, or $t \mu^+$, 
with a mass in the range $200-400$  GeV (so they
decay to an on-shell top).    The production
and decay are calculated by {\sc pythia}\,6.4\cite{pythia}. 
Jets are reconstructed with the anti-$kt$ algorithm
of the {\sc fastjet} package\cite{fastjet}, with $R = .5$.
To estimate a total number 
of surviving events, we assume 1 $fb^{-1}$ of data. 

\subsection{Counting events: $S \to t \mu^+$}
\label{countmu}

Consider first the decay $S \to t \mu^+$. If
one $W$ decays leptonically, this could
 be searched for in   events  with 
$\ell^+ \ell^- + \ETm +$ jets. 
 CMS recently determined
 the $t \bar{t}$ production cross section \cite{CMS}
(with two leptonic $W$s) from such events,
and our 
cuts are  patterned on
this analysis. We expect more
leptons and jets in $S\bar{S}$ production
than in $t \bar{t}$ production, so we impose:
\begin{enumerate}
\item  $\ETm >30$ GeV  
\item  Exactly two  opposite sign
 charged leptons ($e^\pm, \mu^\pm$), with 
 $p_T > 20$, $|\eta| < 2.5$. \\
Or alternatively,  two OS
leptons, with at least one other lepton. 
\item at least four jets of  $p_T >30$ GeV, and  $|\eta| < 2.5$
\end{enumerate}
The CMS analysis has an isolation cut for the leptons;
we instead require that the simulated leptons who
pass $p_T$ cuts be produced in $W$ or $S$ decays
(to avoid high $p_T$ leptons from meson decays).
We
allow all decays to our $W$s in {\sc pythia}.
This means, for instance  that our simulation now
includes  events with 
two leptonic  $W$s, which could pass
cuts if there are  additional QCD jets
(this accounts for $\sim 10\%$  of our events
at $m_S = 200$ GeV).
The fraction of events that survive 
cuts  1, 3 and either of the versions of 2, are respectively  defined as
 $\varepsilon (\ETm, =2OS\ell,4j)$
 and  $\varepsilon (\ETm, > 2OS\ell,4j)$,
and are  given in colomns three and five  of table
 \ref{muepsfin}.
The number of events  at the 7 TeV LHC
with ${\cal L} = 1 fb^{-1}$ of integrated
luminosity 
is estimated  in the fourth
and sixth colomns.

\begin{table}[htp!]
\begin{center}
\begin{tabular}{|c|c|cc|cc|}
\hline
$m_S $&  $\sigma_{prod}$/pb & $\varepsilon(\ETm,=2OS \ell,4j) $ &$ N_=(LQ)$
& $\varepsilon(\ETm,> 2OS \ell, 4j) $ &$ N_>(LQ)$  \\
\hline
200 & 12.5    & .055   &683  &.035 & 438 \\
250 & 3.69   & .095  &352  &.094&346\\
300 & 1.3  & .104  &136 &.116 &151 \\
350 & 0.515   &.109   &56  &.12& 62  \\
400 & 0.224 &  .121& 27 &.129 &29 \\
\hline
\end{tabular}
\caption{The second colomn is the leptquark pair production
cross section at the 7 TeV LHC. The leptoquarks 
decay to $t \mu^\pm$.
  The third colomn  is the
fraction of  events which pass
the cuts of section \ref{countmu} with  exactly a  pair
of opposite sign leptons, and the  fourth colomn is the
estimated number of events 
in  1 $fb^{-1}$ of data. Colomns five and six
are the same, for the  3 or more lepton cut of
section \ref{countmu}.
  \label{muepsfin}
}
\end{center}
\end{table}

We can compare  to the  CMS determination
\cite{CMS}
of the $t \bar{t}$ production cross section,
based on 3.1 $pb^{-1}$ of data from the 7 TeV LHC. 
In the $2 OS \ell$  and $\geq 4$ jet bin, CMS observes one event, 
where $\simeq .75 \to 1.5$ signal events are expected.
From \cite{CMSMC}, it appears that the background
is $\lsim 1/3$ of the signal. We anticipate 
that a 200 GeV leptoquark would contribute
$\sim$ 2 events in the $\geq 4j$ and
exactly 2 OS lepton bin.

The integrated luminosity available now (winter 2011)
is of order 35 $pb^{-1}$, or ten times that used in
the CMS analysis \cite{CMS}.  This suggests that 
$\sigma_{t\bar{t}}$  measurements at the LHC are 
 already sensitive to leptoquarks $S$   with
$BR(S \to t \mu^\pm) = 1$  and $m_S \gsim 200$ GeV. 
Furthermore,  searching for
$\geq 3 \ell$ and $\geq 4$ jets would be more sensitive
to such leptoquarks.

\subsection{Counting events: $S \to t \tau^-$}
\label{counttau}

Consider now the decay 
$S \to t \tau^-$ with $BR(S \to t \tau^-) = 1$. 
This decay  would be more challenging to reconstruct, because the
neutrinos from both $\tau$s and  a leptonic $W$
can contribute to $\ETm$.
We decay 
$S \to t \tau^+$ in {\sc pythia},
and tell {\sc tauola} to flip the sign of   the $\tau$s
from the leptoquarks : $\tau^\pm \to \tau^\mp$,
with helicity assigned as if it were chiral singlet ($\tau_R$). 
Similarly to  the discussion of  $S \to t \tau$ at the Tevatron
(see section \ref{sssttau}), we  attempt to
constrain these leptoquarks at the LHC from
lepton $+$ jets $+ \ETm$ events.  
 However, in our simulation  of $S \to  t \tau$ at
the LHC, unlike that of  section \ref{sssttau}, 
we allow all decays to the $W$s from
the tops. This is because, at the LHC, leptons 
produced in $\tau$ decay can be energetic enough to pass
$p_T$ cuts.

We then impose the following  cuts,
patterned on an  ATLAS \cite{ATLAS} analysis
which extracts the
 $t \bar{t}$ production cross section  from
lepton $+$ jets $+ \ETm$ events:
\begin{enumerate}
\item
  $ \ETm > 25$ GeV, where
all the  neutrinos  are summed  into $\ETm$
\item  at least four
jets with  $p_T >25$ GeV, $|\eta| < 2.5$.
\item  we require at least one   $e^\pm, \mu^\pm$ with 
 $p_T > 20$ GeV, $|\eta| < 2.5$.
To mimic an isolation cut on our simulated leptons,
we then check  that the $e^+, e^-, \mu^+$ and $\mu^-$
with  highest $p_T$s  are  separated from the jets which pass cuts,
by $\Delta R = \sqrt{ (\eta_j - \eta_\ell)^2 + 
(\phi_j - \phi_\ell)^2} \geq .3$.  The leptons failing this
check are rejected. 
 Then, we may
in addition  require
either
exactly one lepton (as in the ATLAS analysis), 
or, at least 7
jets and/or $e^\pm$ and/or $ \mu^\pm$ which pass these cuts.
\end{enumerate}
To take into account both 
 hadronic and leptonic $\tau$ decays, the requirement of
 $\geq 7 $ jets and/or leptons is applied.

 Our  estimates
 of  the fraction of leptoquark
events that would pass these cuts (for
the three possible lepton cuts), and
the number of events in 
1 $fb^{-1}$ of data, are in  table \ref{taueps}.
Notice that  the efficiencies, for finding an
$S \to t \tau^\pm$ leptoquark in such single
lepton events, are higher than the efficiencies
to find $S \to t \mu^\pm$ leptoquarks in
the dilepton events, given in table 
 \ref{muepsfin}.
 This is because there
is always $\ETm$ ($\nu_\tau$s) in the 
$S \to t \tau^\pm$ final state, and  there
is an $e^\pm$ or $\mu^\pm$  approximately
2/3 of the time.  Whereas  requiring   $\ETm$
in the  $S \to t \mu^\pm$  decays,  imposes
a leptonic $W$, which occurs $\sim 29\%$ of the time.

\begin{table}[htp!]
\begin{center}
\begin{tabular}{|c|c|cc|cc|cc|}
\hline
$m_S$ & $ \sigma$/pb & $\varepsilon(\geq 1 \ell ) $
 &$ N(\geq 1 \ell )$ 
&$\varepsilon( 1~\ell)$
&$N( 1~\ell)$
&$\varepsilon(\geq 7 \ell+j)$ 
& $N(\geq 7 \ell+j)$ \\
\hline
200 &12.5 & .160  &2000 & .143  &1788 &  .039 &488 \\
250 &3.69 & .297  &1096  &.241 &889 &.126  & 465  \\
300 &1.30 & .374  &486 &.285 & 370  & .199&259  \\
350 &.515 & .428  &220  &.322 &166 &.234  & 121 \\
400 & .224 & .451  &101 &.328 &74  &.264  & 59 \\
\hline
\end{tabular}
\caption{The second colomn is the leptoquark 
pair production
cross section at the 7 TeV LHC. The leptoquark
decays to $t \tau^\pm$.
 The third, fifth
and seventh  are the
fraction of   events
which  survive the   cuts
of section \ref{counttau}; the
three $\varepsilon$s correspond to
the three different lepton cuts. 
Then the fourth, sixth and eighth colomns
are the estimated number of events passing
cuts in 1 $fb^{-1}$ of data, 
 for $BR(S\to t \tau^-) = 1$.   
 \label{taueps}
}
\end{center}
\end{table}

It is less straightforward to anticipate LHC sensitivity
in this channel. ATLAS \cite{ATLAS} measured
the $t \bar{t}$ production cross section in the single
lepton $+$ jets $+ \ETm$ channel, with 2.8 $pb^{-1}$ of data.
In the bin containing  $\geq 4 $ jets  of which  two are
tagged as $b$s, they observe 37 events, expected 30
from $t \bar{t}$,  and estimate the background as
$12.2 \pm 3.9$.  The ATLAS $b$ tagging efficiency varies
($40 -60 \%$ for$ E_j: 25 \to 85 $GeV); if we assume
that $\sim 50\%$ of the  $b$s from leptoquarks are tagged,
one can guess from  table \ref{taueps}
that a $m_S = 200$ GeV leptoquark, with
$BR(S \to t \tau^-) = 1$, could contribute  $\sim  1-2$ 
events to this bin. Explicitely counting events with 
extra jets (beyond
the four expected from
$t \bar{t}$),  or events with  extra  leptons 
(ATLAS required one and  only one), could
improve the sensitivity to leptoquarks
decaying to $t \tau^-$.


\section{Summary}
\label{}

Like the Higgs, a  scalar leptoquark  is a   boson that couples to  two
fermions.  Since the  quark Yukawa couplings to the Higgs 
are hierarchical, and  flavour physics in the quark
sector follows Standard Model expectations,   one can
anticipate that leptoquarks, like
the Higgs, interact preferentially with third generation quarks.
This also arises in
several models.   However, expectations for 
leptoquark couplings to leptons    are less straightforward to 
extract from lepton mass matrices. The charged leptons
are hierarchical, so one could imagine that leptoquarks
should preferentially decay to $t \tau^\pm$. On the other
hand, the neutrino sector is comparatively democratic,
suggesting that leptoquarks could decay to $t$ and any
lepton.  The $te^\pm$ and $t \mu^\pm$ final states
could be interesting search channels for the early LHC.

This paper studied possible bounds on
leptoquarks $S$, with a mass in
the range $100$ GeV $ < m_S < 400$ GeV, 
which are pair-produced via their
strong interactions, and 
decay to a  top quark
(and only the top; no $b, c...$),
and a charged lepton $L = e, \mu$ or $\tau$.
We expect the  Tevatron, with its high
luminosity, 
to be sensitive to  the range  $m_S < m_t$,
where the leptoquarks decay to
the three-body final state $b W^\pm L^\mp$. 
 Leptoquarks
with   $m_S > m_t$  could be found at the LHC.
The range   $m_S  \simeq m_t \pm 15 $ GeV
appears difficult: the soft  leptons  in the final
state may not pass $p_T$ cuts,  so the $S \bar{S}$ final
state becomes difficult to distinguish from
$t \bar{t}$. Recall that at $m_S \sim m_t$,  the
$S\bar{S}$ production cross section is 
$\sim 1/10$ of the $t \bar{t}$ production
cross section.

We estimated the  number of leptoquark-induced
events containing  lepton(s) plus jets, using
{\sc pythia}\,6.4, {\sc tauola}, and  the anti-$k_t$ jet
algorithm of {\sc fastjet}.  We include no detector
simulation. We consider separately the sensitivities
to leptoquarks
which decay to $t \tau^\pm$, or $t \mu^\pm$,
assuming a branching ratio of 1 in each case. 
We  further  assume that our estimates for the $t \mu^\pm$
final state could apply to  leptoquarks
decaying to  $t e^\pm$.

Our results suggest that
current determinations of the $t\bar{t}$ production cross section,
both from the Tevatron and the LHC, could constrain
a leptoquark with $BR( S \to t \mu^\pm) = 1$, and a mass
of order  $100 \to 250$ GeV. 
At the Tevatron, 
D0 determines the $t \bar{t}$
 production cross section  $\sigma_{t \bar{t}}$
 from the final state
$\ell_i^\pm \ell_j^\mp + \ETm + \geq 3 $ jets,
using  4.3 $fb^{-1}$ of data.
We  estimate
that these results could 
exclude
$$
m_S \lsim 160  ~{\rm GeV} ~~~~~~~
 {\rm for} ~~~BR(S \to t~ \ell^\pm) = 1, ~~~~~ \ell \in \{ e \mu \}
$$   
At the LHC, CMS 
obtained  $\sigma_{t \bar{t}}$
 from  events containing 
$\ell_i^\pm \ell_j^\mp + \ETm + \geq 2 $ jets.
It is possible
that leptoquarks with  $m_S \gsim 200$ GeV,
could have contributed a few events to the
scantily populated (1 event)  $\geq 4 $ jet bin.

Leptoquarks with  $BR(S \to t \tau^\pm$)
would contribute  to the  
 the final state
$\ell + \ETm + \geq 4 $ jets
(which is  used to determine 
 $\sigma_{t \bar{t}}$),
if at least  one of the $W$s
or $\tau$s decays leptonically.
  A significant
fraction of the  leptoquark
events should have more than four jets,
but the available data sets present
a single  $\geq 4 $ jet bin, so
we cannot  profit from this property.
We estimate that  a D0 analysis could exclude  
$$
m_S \lsim 160  ~{\rm GeV} ~~~~~~~
 {\rm for} ~~~BR(S \to t \tau^\pm) = 1
$$ 
ATLAS also  obtained  $\sigma_{t \bar{t}}$
from  events with 
$\ell + \ETm + \geq 4 $ jets, 
 but leptoquarks
with $m_S \sim 200$ GeV 
would be consistent with the backgrounds.

The current integrated  luminosity of
the  LHC is 
significantly larger than that  used
 in the analyses we compared to  \cite{ATLAS,CMS}.
So we anticipate that  the  winter 2011
determinations of   $\sigma_{t \bar{t}} $ 
at the LHC should have some sensitivity to 
the leptoquarks discussed here. 
However,  $S \bar{S}$
production  followed by
$S \to t L^\pm$,  should usually
give  a final state with  more leptons  and/or jets than
$t \bar{t}$ production. This means that  
 analyses   counting
 additional  leptons and/or jets,
 could 
have improved sensitivity.

\section*{Acknowledgements}
We thank A Aparici, C Biscarat, G Grenier,
S Perries sand V Sanz for useful discussions.

\end{document}